\documentclass[sigconf]{acmart}
\usepackage{hyperref}
\usepackage{epsfig}
\usepackage{graphicx}
\usepackage{amsmath}
\usepackage{amssymb}
\usepackage{helvet}  %Required
\usepackage{courier}  %Required
\usepackage{url}  %Required
\usepackage{xcolor}
\usepackage{multirow}
\usepackage{subcaption}
\usepackage{array}
\usepackage[export]{adjustbox}

\usepackage{balance}
\newcolumntype{P}[1]{>{\centering\arraybackslash}p{#1}}

\def\BibTeX{{\rm B\kern-.05em{\sc i\kern-.025em b}\kern-.08emT\kern-.1667em\lower.7ex\hbox{E}\kern-.125emX}}
    
\copyrightyear{2020}
\acmYear{2020}
\setcopyright{acmlicensed}
\acmConference[ICMR '20]{Proceedings of the 2020 International Conference on Multimedia Retrieval}{June 8--11, 2020}{Dublin, Ireland}
\acmBooktitle{Proceedings of the 2020 International Conference on Multimedia Retrieval (ICMR '20), June 8--11, 2020, Dublin, Ireland}
\acmPrice{15.00}
\acmDOI{10.1145/3372278.3390695}
\acmISBN{978-1-4503-7087-5/20/06}

\begin{document}

\fancyhead{}
  % do not delete this code.

% The "title" command has an optional parameter, allowing the author to define a "short title" to be used in page headers.
\title{Query-controllable Video Summarization}

% The "author" command and its associated commands are used to define the authors and their affiliations.
% Of note is the shared affiliation of the first two authors, and the "authornote" and "authornotemark" commands
% used to denote shared contribution to the research.
% \author{Jia-Hong Huang}
% \email{j.huang@uva.nl}
% \affiliation{%
%   \institution{University of Amsterdam}

% \author{Marcel Worring}
% \affiliation{%
% \email{m.worring@uva.nl}

% \author{Jia-Hong Huang}
% \affiliation{%
%   \institution{University of Amsterdam}

% \affiliation{%
 
% \affiliation{%

% \affiliation{%

\author{Jia-Hong Huang}
\affiliation{
  \institution{University of Amsterdam, Amsterdam, Netherlands}
%   \city{Amsterdam Netherlands}
%   \country{Netherlands}
  }
\email{j.huang@uva.nl}

\author{Marcel Worring}
\affiliation{
  \institution{University of Amsterdam, Amsterdam, Netherlands}
%   \city{Amsterdam, Netherlands}
%   \country{Netherlands}
  }
\email{m.worring@uva.nl}

%
% By default, the full list of authors will be used in the page headers. Often, this list is too long, and will overlap
% other information printed in the page headers. This command allows the author to define a more concise list
% of authors' names for this purpose.
\renewcommand{\shortauthors}{Trovato and Tobin, et al.}

%
% The abstract is a short summary of the work to be presented in the article.
\begin{abstract}
When video collections become huge, how to explore both within and across videos efficiently is challenging. Video summarization is one of the ways to tackle this issue. Traditional summarization approaches limit the effectiveness of video exploration because they only generate one fixed video summary for a given input video independent of the information need of the user. In this work, we introduce a method which takes a text-based query as input and generates a video summary corresponding to it. We do so by modeling video summarization as a supervised learning problem and propose an end-to-end deep learning based method for query-controllable video summarization to generate a query-dependent video summary. Our proposed method consists of a video summary controller, video summary generator, and video summary output module. To foster the research of query-controllable video summarization and conduct our experiments, we introduce a dataset that contains frame-based relevance score labels. Based on our experimental result, it shows that the text-based query helps control the video summary. It also shows the text-based query improves our model performance. 
\textbf{\href{https://github.com/Jhhuangkay/Query-controllable-Video-Summarization}{\scriptsize{https://github.com/Jhhuangkay/Query-controllable-Video-Summarization}}.}

\end{abstract}




%
% Keywords. The author(s) should pick words that accurately describe the work being
% presented. Separate the keywords with commas.

% \keywords{Adaptive Summarization; Video Summarization; Video Summary Controller; Video Summary Generator; Multi-modal Feature Fusion; Text-based Input Query}

%
% A "teaser" image appears between the author and affiliation information and the body 
% of the document, and typically spans the page. 
%   \Description{Enjoying the baseball game from the third-base seats. Ichiro Suzuki preparing to bat.}

%
% This command processes the author and affiliation and title information and builds
% the first part of the formatted document.

\maketitle

\section{Introduction} 
Video data is now ubiquitous in our daily life. Most of the raw videos are too long and are containing redundant content. As a consequence, the amount of video data people have to watch is overwhelming. This raises new challenges in efficiently exploring both within and across videos. Video summarization \cite{gong2014diverse,zhang2016summary,zhang2019dtr,zhou2018deep,vasudevan2017query} helps people explore a video efficiently by capturing the essence of the video. Learning what is essential depends on the information need of the user. Yet, traditional video summarization methods, such as \cite{ngo2003automatic,song2015tvsum,de2011vsumm,chu2015video,kang2006space,lee2012discovering,gygli2014creating,yang2019causal}, generate one fixed video summary for a given input video. Hence, they either create a video capturing all possible information needs and therefore yield limited reduction in time, or they lose essential information for specific needs. Having a fixed summary limits the effectiveness of video exploration. 

To make the video exploration more effective and efficient, we will need a new specialized method that, steered by the information need of the user, is capable of generating various video summaries for a given video. We call this query-controllable video summarization. There are two main features of query-controllable video summarization which complicate this task when compared to the well studied domain of conventional video summarization. One example is that query-controllable video summarization has a text-based query input and a video input, where the conventional video summarization only has a video input. So, in query-controllable video summarization, we need to model the implicit relations or interactions between the input query and video. Evaluating the generated video summaries is another challenge. Previously, researchers usually conduct human expert evaluation based on predefined rules or showing human experts two different video summaries and asking them to select the better one \cite{over2008trecvid,lu2013story,lee2012discovering}. Human expert based evaluation methods for this task are problematic, as these methods are expensive and time-consuming because they rely on the judgments of humans for each evaluation \cite{gygli2014creating}. In this paper, we prefer to conduct automatic evaluation which is more efficient.

\begin{figure}[t!]
\begin{center}
\includegraphics[width=1.0\linewidth]{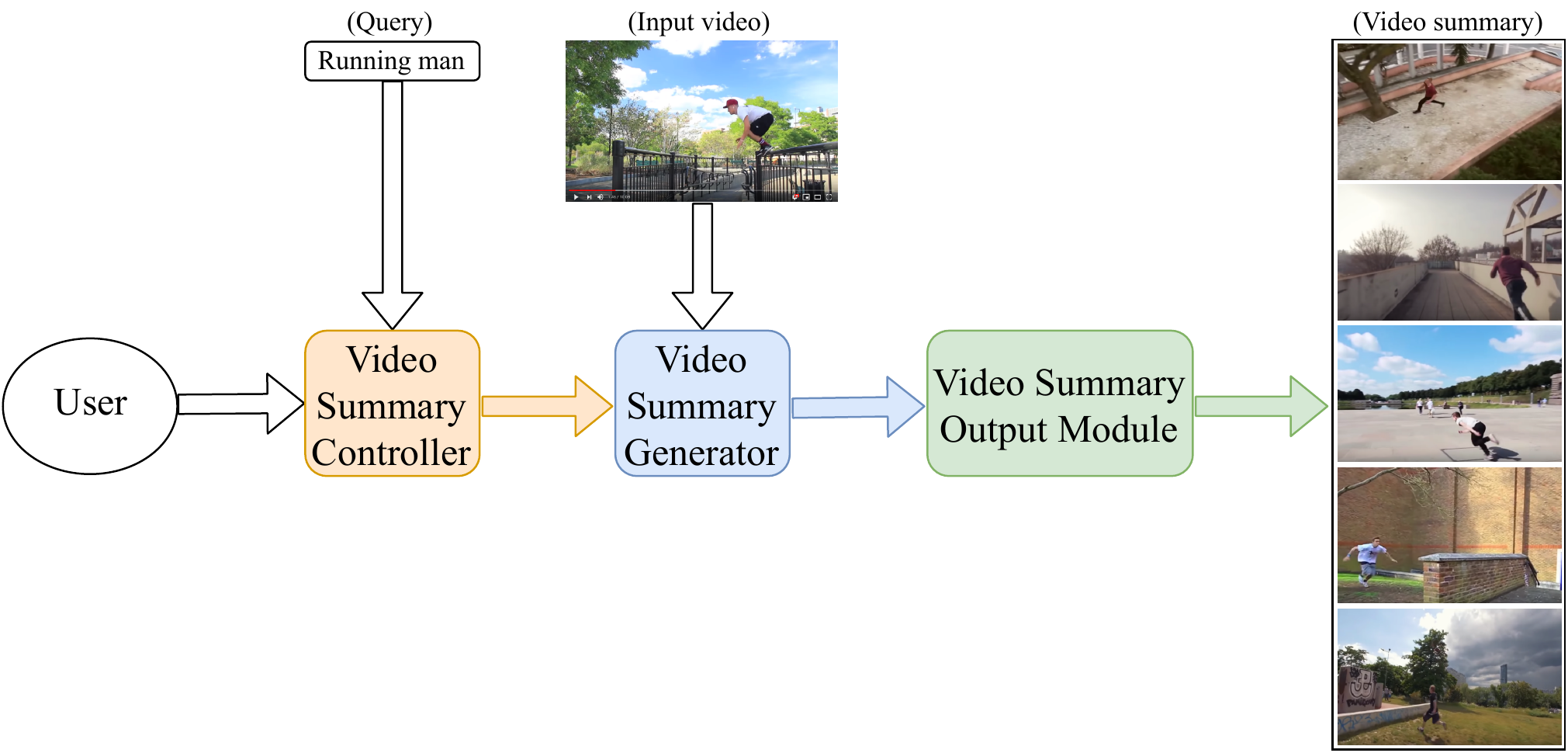}
\end{center}
\vspace{-0.40cm}
   \caption{This figure shows the main idea of the query-controllable video summarization. A user gives a text-based query, which represents the desired video summary, to control a video summary generator to create a video summary based on the query. Note that the input of the proposed method is both query and video. If two different queries are given with the same video input, two different query-dependent video summaries will be generated. It is different from the traditional video summarization which only has a video input. For details of the video summary controller and video summary generator, please refer to Figure \ref{fig:figure6} and \ref{fig:figure1}.}
\vspace{-0.30cm}
\label{fig:figure100}
\end{figure}

There are different solutions to model the video summarization problem, both supervised \cite{gong2014diverse,gygli2014creating,gygli2015video,li2018local,rochan2018video,zhang2016summary,zhang2016video,zhang2018retrospective,zhang2019dtr,zhou2018deep} and unsupervised \cite{de2011vsumm,chu2015video,kang2006space,lee2012discovering,liu2002optimization, ma2002user,panda2017collaborative,ngo2003automatic,potapov2014category,rochan2019video,zhao2014quasi,liu2019interpretable}. For the automatic evaluation, in this paper, we model video summarization as a supervised learning problem. Then, we propose an end-to-end deep learning based model, illustrated in Figure \ref{fig:figure100}, to generate video summaries depending on the text-based input queries. To the best of our knowledge, we are the first proposing an end-to-end deep model for query-controllable video summarization.
Note that a non-end-to-end method for query-controllable video summarization needs many preprocessing steps, and in practice, it will reduce the efficiency of video exploration. 
% \textcolor{blue}{start to explain the proposed method.}
The model consists of a video summary controller, a video summary generator, and a video summary output module. The controller uses text, such as words, phrases, or sentences, to describe the desired video summary. Then, the generator creates a video summary based on the implicit relationship between the text-based description and the input video. The video summary output module then outputs a video summary based on a relevance score prediction vector.

To build a query-controllable system for videos, using multiple information sources requires dedicated datasets and evaluation methods. We undertake such a study in this work. Starting from an existing dataset \cite{vasudevan2017query}, we establish a new video dataset on which to build our models. With such a dataset in place, it is the right time to research how to exploit deep learning-based models with textual query inputs which are to steer the output. 

As mentioned above, our proposed model takes a text-based query and video as input, so how to effectively fuse the multi-modal features with minimum loss of information is one of the technical problems. It has been shown that, in similar multi-modal contexts, the performance of models can decrease if feature fusion methods are improper; how to solve this issue, in general, remains an open question \cite{ben2017mutan,fukui2016multimodal}. So, we not only conduct experiments to see how the input query affects our model performance, but also conduct experiments to see how the commonly used feature fusion methods affect our model performance. To foster the research community of video summarization, we intend to publish the dataset and code.

We demonstrate experimental results of the proposed end-to-end deep model and show that our proposed method is capable of creating query-dependent and controllable video summaries for given videos. Our experimental result shows that the text-based input query helps control the video summary. It also shows that the text-based input query improves the performance of our summarization model, +$5.83$\% in the sense of accuracy.

\noindent\textbf{Contributions.}
% \vspace{-4pt}
\begin{itemize}
    \item We propose a new end-to-end deep learning based model for a text-visual embedding space for query-controllable video summarization.
    % Based on our experimental results, the proposed method successfully improves the traditional treatment procedure of retinal diseases.
    % We present a patient-centered method for deep learning-aid medical prediagnosis on ophthalmic disease. 
    
    \item We conduct detailed experiments to show that the text-based query not only helps control the video summary but also improves the performance of the proposed end-to-end model.
    % We present a new clinical labels collection, DeepEyeNet, we present a large medical label collection included vision and text features.
    \item  We introduce a query-video pair based dataset, based on the dataset proposed in \cite{vasudevan2017query}, for a query-controllable video summarization task. The dataset contains 190 videos and corresponding frame-based relevance score annotations.

    % \item Finally, to further understand the deep networks in retinal disease classification, we visualize the learned features by heat maps and compare to the other our proposed retinal images collection, manually labeled by ophthalmologists. The visualization result shows that our model focus on the similar part of retinal image as ophthalmologists do when they try to judge the retinal image. Besides, we propose a couple of methods or baselines trained on our DEN.
    % Finally, we visualize the learned features inside the DNNs model by heat maps. The visualization helps in understanding the medical comprehensibility inside our DNNs model.(the results show that our model focus on the same part as eye dector when they do the decision.)
    
    % \item \textcolor{red}{Finally, the traditional text evaluation metrics are not suitable in evaluating the clinical description, so we propose a new metric, causal scores, for calculating the similarity between two clinical descriptions by text-graph model.}
    % We present causal scores for calculating the similarity between two sentences by text-graph model.
\end{itemize}

\section{Related Work}
In this section, we discuss related work in terms of different methods and different datasets. We first discuss the two main types of methods of video summarization, i.e., supervised and unsupervised. Then, we review the commonly used video summarization datasets. 

% \vspace{+0.37cm}
\noindent\textbf{2.1 Unsupervised Video Summarization}

Unsupervised approaches for video summarization, \cite{de2011vsumm,chu2015video,kang2006space,lee2012discovering,lu2013story,liu2002optimization, ma2002user,panda2017collaborative,ngo2003automatic,song2015tvsum,potapov2014category,rochan2019video,zhao2014quasi,vasudevan2017query}, usually exploit hand-crafted heuristics to satisfy some properties, such as interestingness, representativeness, and diversity. In \cite{de2011vsumm}, the authors propose a method, based on color feature extraction from video frames and k-means clustering. The authors of \cite{chu2015video} observe that the key visual concepts usually appear repeatedly across videos with the same topic. So, they propose to create a summarized video by finding shots that co-occur most frequently across videos. They develop a Maximal Biclique Finding (MBF) algorithm to find sparsely co-occurring patterns. In \cite{kang2006space}, the authors introduce a space-time video summarization method to extract the visually informative space-time portions of input videos and analyze the distribution of the spatial and temporal information in a video, simultaneously. The authors of \cite{lee2012discovering} present a video summarization method for egocentric camera data. They develop region cues, e.g., the nearness to hands, gaze, and frequency of occurrence, in egocentric video and learn a model to predict the relative importance of a new region based on these cues. 
% In \cite{liu2002optimization}, the authors model video summarization as a problem of keyframes selection and propose a new optimization-based method to tackle it. Note that they define keyframes to be a temporally ordered subsequence of a given video sequence. 
The authors of \cite{lu2013story} present a video summarization method to discover the story of a given egocentric video. Their proposed method is capable of selecting a short chain of sub shots of video which depicting the essential events. In \cite{ma2002user}, based on the modeling of the viewer’s attention, the authors propose a generic framework of video summarization. The framework, without the fully semantic content understanding of a given video, eliminates the needs of complicated heuristic rules in video summarization task and takes advantage of computational attention models. In \cite{ngo2003automatic}, the authors introduce a unified method for video summarization based on the analysis of structures and highlights of the video. Their approach emphasizes the content balance and perceptual quality of a video summary at the same time. They also incorporate a normalized cut algorithm to partition a video into clusters and a motion attention model based on human perception to compute the perceptual quality of clusters and shots. The authors of \cite{panda2017collaborative} develop a new method to extract a video summary that captures important particularities arising in a given video and generalities identified from a set of given videos simultaneously. 
% In \cite{potapov2014category}, the authors propose a category-specific video summarization method to generate short and high informative video summaries. Their approach exploits a support vector machine (SVM) classifier to assign importance scores to each semantically-consistent segment created by performing a temporal segmentation. The authors of \cite{song2015tvsum} observe that the title of a video is usually chosen to best describe its main topic. Then, images related to the title serve as a proxy for key visual concepts of the topic. So, they propose an unsupervised video summarization framework based on title-based image search results to find visually important shots. 
The authors of \cite{zhao2014quasi} introduce a method to learn a dictionary from the given video using group sparse coding. A video summary is then generated by combining segments that cannot be reconstructed sparsely using the dictionary. In \cite{rochan2019video}, the authors propose a new formulation to perform video summarization from unpaired data. Their model aims to learn mapping, from a set of raw videos to a set of video summaries, such that the distribution of the generated video summary is similar to the distribution of the set of video summaries with the help of an adversarial objective. Also, they enforce a diversity constraint on the mapping to ensure the generated video summaries are diverse enough visually. In \cite{zhou2018deep}, the authors introduce an end-to-end reinforcement learning-based framework to train their video summarization model. The framework incorporates a new reward function to jointly account for diversity and representativeness of generated video summaries. The design of the reward function makes it not rely on user interactions or labels at all. The authors of \cite{vasudevan2017query} propose a more optimization-based method with a specific objective for query-adaptive video summarization. Since the method is not end-to-end, it will need many preprocessing steps before generating a video summary. So, in practice, it is inconvenient and not efficient for video exploration. This motivates us to propose an end-to-end deep learning based method. Although many existing works model video summarization task as an unsupervised problem and propose their methods to tackle it, in general, the performance of the unsupervised approach is worse than the supervised one. 
% So, in this paper, we prefer to model the task as a supervised learning problem.

% \textcolor{blue}{Some summarization methods also provide weak supervision through additional cues such as web images/videos [4, 13, 14, 33] and video category information [28, 30] to improve the performance}

% \vspace{+0.1cm}
\noindent\textbf{2.2 Supervised Video Summarization}

The other type of approach for video summarization task is supervised, \cite{gong2014diverse,gygli2014creating,gygli2015video,li2018local,rochan2018video,sharghi2018improving,zhang2016summary,zhang2016video,zhang2018retrospective,zhang2018query,zhang2019dtr,zhou2018deep}. The learning of these methods is supervised by human expert labeled data, i.e., ground truth video summaries. The authors of \cite{gong2014diverse} treat video summarization as a supervised subset selection task. They propose a probabilistic model for selecting a diverse sequential subset, called the sequential determinantal point process (SeqDPP). Note that the standard DPP treats video frames as randomly permutable elements. The SeqDPP heeds the inherent sequential structures in video data, so it not only overcomes the deficiency of the standard DPP, but also retains the power of modeling diverse subsets, which is essential for video summarization. The authors of \cite{gygli2014creating} introduce a new video summarization method and focus on user videos containing a set of interesting events. The method starts by segmenting a given video based on a superframe segmentation, tailored to raw videos. Then, according to the estimation score of visual interestingness per superframe, by using a set of low-level, mid-level, and high-level features, the method picks an optimal subset of superframes to generate a video summary. In \cite{gygli2015video}, the authors propose a new model to learn the importance score of global characteristics of a video summary. The models, jointly optimized for multiple objectives, is capable of generating high-quality video summaries. In \cite{li2018local}, the authors introduce a new probabilistic model, built upon SeqDPP, to tackle video summarization problem. The period of a video segment, where the local diversity is imposed, can be dynamically controlled by the model. To get a well-trained summarization model, the authors develop a reinforcement learning algorithm to train the proposed model. The authors of \cite{rochan2018video} formulate video summarization as a sequence labeling problem. They propose fully convolutional sequence models to tackle the video summarization task. First, they establish a novel connection between video summarization and semantic segmentation. Second, the adapted popular semantic segmentation networks are used to generate video summaries. In \cite{sharghi2018improving}, the authors propose an improved sequential determinantal point process (SeqDPP) model. In terms of modeling, a new probabilistic distribution is designed to make, when it is integrated into SeqDPP, the resulting model accepts the input of a user, about the intended length of the video summary. In terms of learning, a large-margin algorithm is proposed to address the problem of exposure bias in SeqDPP. In \cite{zhang2016summary}, the authors propose a subset selection method that leverages supervision in the form of human-created video summaries to perform keyframe-based video summarization. The main idea of this method is nonparametrically transferring structures of summaries from annotated videos to unseen testing videos. Also, the authors generalize the proposed method to sub-shot-based video summarization. The authors of \cite{zhang2016video} cast a video summarization task as a structured prediction problem on sequential data. Then, they propose a new supervised learning technique, incorporating Long Short-Term Memory (LSTM), to model the variable-range dependencies entailed in the task. Also, they exploit domain adaptation techniques, based on the auxiliary annotated video datasets, to improve the quality of the video summary. In \cite{zhang2018retrospective}, the authors propose a sequence-to-sequence learning model to tackle the video summarization problem. To complement the discriminative losses with another loss, such as measuring whether the generated video summary preserves the same information as the original video, they propose to augment standard sequence learning models with a retrospective encoder that embeds the predicted video summary into an abstract semantic space. Then, the embedding is compared to the original video's embedding in the same space. 
% The authors of \cite{zhang2018query} introduce a query-conditioned three-player generative adversarial network to tackle video summarization task. The joint representation of the query and video content is learned by a generator. The discriminator takes three pairs of query-conditioned video summaries as the input to discriminate the real video summary from a predicted and a random one.  
The authors of \cite{zhang2019dtr} introduce a novel dilated temporal relational generative adversarial network (DTR-GAN) to tackle the frame-based video summarization. DTR-GAN exploits an adversarial manner with a three-player loss to learn a dilated temporal relational generator and discriminator. The authors introduce a new dilated temporal relational unit to enhance the capturing of temporal representation, and then the generator creates keyframes based on the unit. The Supervised methods are capable of learning useful cues, which are hard to capture with hand-crafted heuristics, from ground truth video summaries. So, they usually outperform the unsupervised models. That is the reason why we prefer to  
model video summarization as a supervised learning problem in this paper.

\begin{table*}[t!]
    \caption{Summary of commonly used video summarization datasets. Based on this table, we find that the proposed dataset is much larger than the other datasets, and it contains two types of input modalities including video and text. The other two video summarization dataset only contains video data, and the dataset size is not large. So, the proposed dataset is unique. Relevance scores in this work and important scores from TVSum are different, referring to \textit{``Crowd-sourced Annotation''} subsection.}
    \vspace{-0.4cm}
\begin{center}
\scalebox{1.0}{
%\scalebox{0.412}{
    \begin{tabular}{|P{3.0cm}|P{3.205cm}|P{3.205cm}|P{3.205cm}|P{3.0cm}|}
    \hline
    \textbf{Name of Dataset} & \textbf{Annotation Type} & \textbf{Content} & \textbf{Number of Videos} & \textbf{Input Modality}\\ \hline
    %SVM & &\\ \hline
    SumMe \cite{gygli2014creating} & Interval-based shot and frame-level scores & User videos & 25 & Video\\ \hline
    TVSum \cite{song2015tvsum} & Frame-based important scores & YouTube videos & 50 & Video\\ \hline
    \textbf{Ours based on \cite{vasudevan2017query}} & \textbf{Frame-based relevance scores} & \textbf{YouTube videos} & \textbf{190} & \textbf{Video, Text}\\ \hline 
    
    \end{tabular}}
        % \vspace{-0.6cm}

    % \vspace{-0.1cm}
    \label{table:table103}
\vspace{-0.3cm}
\end{center}
\end{table*}

% \vspace{+0.1cm}
\noindent\textbf{2.3 Video Summarization Dataset Comparison}

In this section, we shortly introduce a commonly used video summarization dataset, \cite{song2015tvsum,gygli2014creating}, and do some comparison with the dataset used in this paper. To tackle the video summarization task, the authors of \cite{song2015tvsum} propose a dataset, named TVSum. It contains 50 videos, with 10 categories, and the corresponding shot-level importance scores obtained via crowdsourcing. The 10 categories are selected from the TRECVid Multimedia Event Detection (MED) task \cite{smeaton2006evaluation}, and the 50 videos, five per category, are collected from YouTube by using the names of categories as search queries. From the search results, videos are chosen based on the following criteria: (i) the selected video should contain more than a single shot; (ii) the title of video is descriptive of the visual topic in the video; (iii) under the Creative Commons license; (iv) the duration of video is around 2 and 10 minutes. The authors exploit Amazon Mechanical Turk (AMT) to collect 1,000 responses, 20 per video, and these responses are treated as gold standard labels \cite{gygli2014creating,khosla2013large,potapov2014category,huang2017vqabq,yang2018novel,huang2018robustness,hu2019silco,yang2018auto,liu2018synthesizing,huang2019assessing,huang2017robustness}. A participant from AMT is asked to (1) read the title of video first, simulating a typical scenario of online video browsing; (ii) watch the whole video in a single take; (iii) provide an importance score to each of uniform length shots for the whole video, denoting from 1 (not important) to 5 (very important). The audio is muted to ensure the important scores are only based on visual stimuli. According to the authors' experience, a two-second shot length is appropriate for capturing local context with good visual coherence. In \cite{gygli2014creating}, the authors introduce another video summarization benchmark, called SumMe, consisting of 25 videos, covering holidays, events and sports. The length of video ranges from 1 to 6 minutes and each video is summarized by 15 to 18 different people. The authors asked 19 males and 22 females to participate in making the dataset. Given a video, participants are asked to produce a video summary containing most of the important content in the video. They are allowed to watch, cut, and edit a video by using a simple interface. The length of a video summary is required to range from 5\% to 15\% of the original video length. That is to ensure the input video is indeed summarized rather than being shortened slightly. The videos are shown randomly and the audio is muted to ensure the generated video summaries are only based on visual stimuli. Regarding the evaluation of video summarization approaches, previously, researchers conduct the human expert evaluation in one of the following ways: i) based on a set of predefined criteria, such as the degree of redundancy, counting the inclusion of predefined important content, summary duration, and so on \cite{over2008trecvid}. ii) showing human experts two different video summaries and asking them to select the better one \cite{lu2013story,lee2012discovering}. The authors of \cite{gygli2014creating} claim that the above human expert evaluation methods are problematic, as these methods are expensive and time-consuming because they rely on judges of human for each evaluation. For example, in \cite{lu2013story}, the evaluation of the method requires one full week of human labor. Both of the human expert evaluation methods help to tell which video summary is better than another but fail to show what a good video summary should look like. So, the authors of \cite{gygli2014creating} do not exploit the above approaches. Instead, they let a set of participants create their video summaries and collect multiple video summaries for each video. The reason is that there is no true answer for correct video summarization, but rather multiple possible ways. With these human expert video summaries, they can compare any summarization method which creates an automatic video summary in a repeatable and efficient way. In \cite{khosla2013large,de2011vsumm}, such automatic versus human comparison has already been used successfully for keyframes. Also, the authors of \cite{khosla2013large} show that comparing automatic keyframe-based summaries to human keyframe-based selections yields ratings that are comparable to letting humans directly judge the automatic video summaries. Both TVSum and SumMe datasets allow the automatic evaluation of video summarization approaches. In this paper, we also establish a dataset, based on \cite{vasudevan2017query}, with automatic evaluation for our query-controllable video summarization task. The proposed dataset contains 190 videos with frame-level relevance score annotations. For convenience, we summarize the above existing datasets and comparison with our dataset in Table \ref{table:table103}.

\section{Dataset Introduction and Analysis}
In this section, we start to describe and analyze our proposed dataset for query-controllable video summarization in terms of types of videos, video labels, and some statistics of the dataset. Note that although the dataset from \cite{vasudevan2017query} partially matches our research purpose and is publicly available, we discover that the specification, such as annotations and amount of videos, of the published dataset are different from the one mentioned in \cite{vasudevan2017query}. Also, some parts of the published dataset are not available anymore. We base our evaluation on the dataset published in \cite{vasudevan2017query}. So, we will first describe the process they have used to create the dataset and from there indicate what changes needed to make it suitable for our purpose.

% Since the authors of \cite{vasudevan2017query} publish their raw dataset, link-based and frame by frame, and do not provide code to download the dataset, we implement the crawling code to download the dataset frame by frame. After that, we spend quite some time to reorganize and clean the dataset and form a new dataset based on it.
% To the best of our knowledge, query-controllable video summarization is a relatively unexplored and challenging domain. So, in this case, we cannot find a proper dataset for our research purpose. We decide to make a new one based on the existing dataset \cite{vasudevan2017query}. To foster the research of query-controllable video summarization, we intend to publish our complete dataset.

\begin{figure}[t]
\begin{center}
\includegraphics[width=1.0\linewidth]{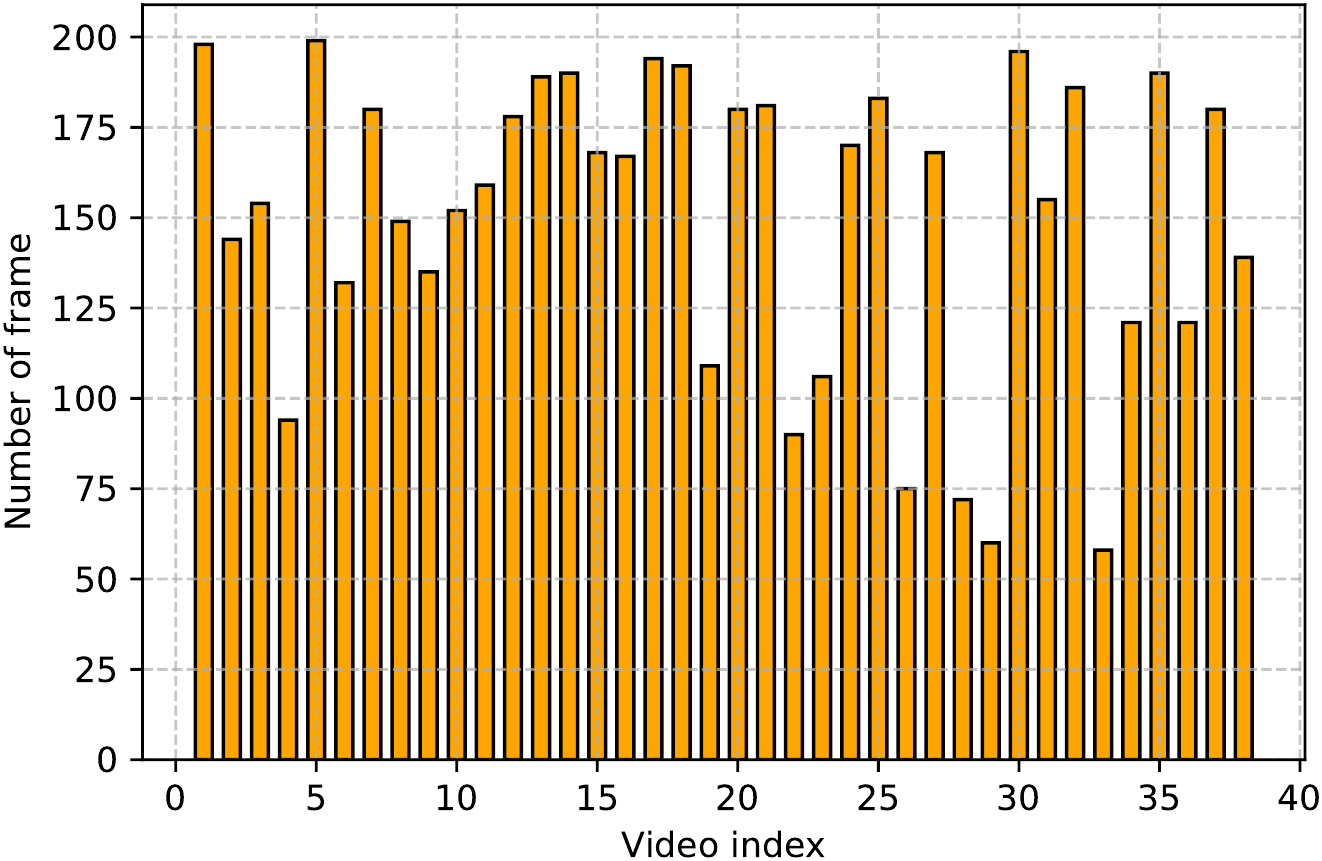}
\end{center}
\vspace{-0.40cm}
   \caption{This figure shows the original number of frames of each video. The x-axis denotes the video index and the y-axis indicates the number of frames. Note that, for convenience, we make all the videos have the same number of frames when we develop our proposed method.
   }
\vspace{-0.30cm}
\label{fig:figure2}
\end{figure}

% \vspace{+0.37cm}
\noindent\textbf{3.1 Setup}

Since our dataset is based on \cite{vasudevan2017query}, the rules from \cite{vasudevan2017query} for the dataset collection are similar. The proposed dataset consists of 190 videos and each video is retrieved based on a given text-based query. Then, according to \cite{vasudevan2017query}, the authors use Amazon Mechanical Turk (AMT) to annotate the video frames with the labels of text-based query relevance scores. The labels in this dataset are used similarly in the MediaEval diverse social images challenge \cite{ionescu2014retrieving}. The purpose of human expert annotated labels in the proposed dataset is to automatically evaluate the methods for generating relevant and diverse video summaries. In the proposed dataset based on \cite{vasudevan2017query}, the representative samples of queries and videos are collected based on the following procedure: The seed queries, with 22 different categories, are selected from the top YouTube queries between 2008 and 2016. Typically, since these queries are generic concepts and short, the YouTube auto-complete function is exploited to obtain more realistic and longer queries, e.g., ``ariana grande focus instrumental'' and ``ark survival evolved dragon''. For each query, the top video result with a duration of 2 to 3 minutes is collected. The following task is set up on AMT by \cite{vasudevan2017query} for video annotations. All of the 190 videos are sampled at one frame per second (fps). Then, an AMT worker is asked to annotate every frame with its relevance with respect to the given text-based query. The answer candidates are ``Very Good'', ``Good'', ``Not good'', and ``Bad'', where ``Bad'' denotes the frame is not relevant and low-quality, such as bad contrast, blurred, and so on. To reduce the subjectivity of labels, every video in the proposed dataset is annotated by at least 5 different AMT workers. Additionally, a qualification task is defined to ensure high-quality annotations. The results are manually reviewed to make sure the workers provide annotations with good quality. After workers pass this task, then they are allowed to take further assignments. Since our maximum number of frames of video is 199, to make all the videos have the same number of frames, 199, we repeat frames starting from the first frame until reaching 199 frames in total for each video, similar to \cite{sigurdsson2017asynchronous}. Figure \ref{fig:figure2} shows the original number of frames of each video.

\begin{figure}[t!]
\begin{center}
\includegraphics[width=1.0\linewidth]{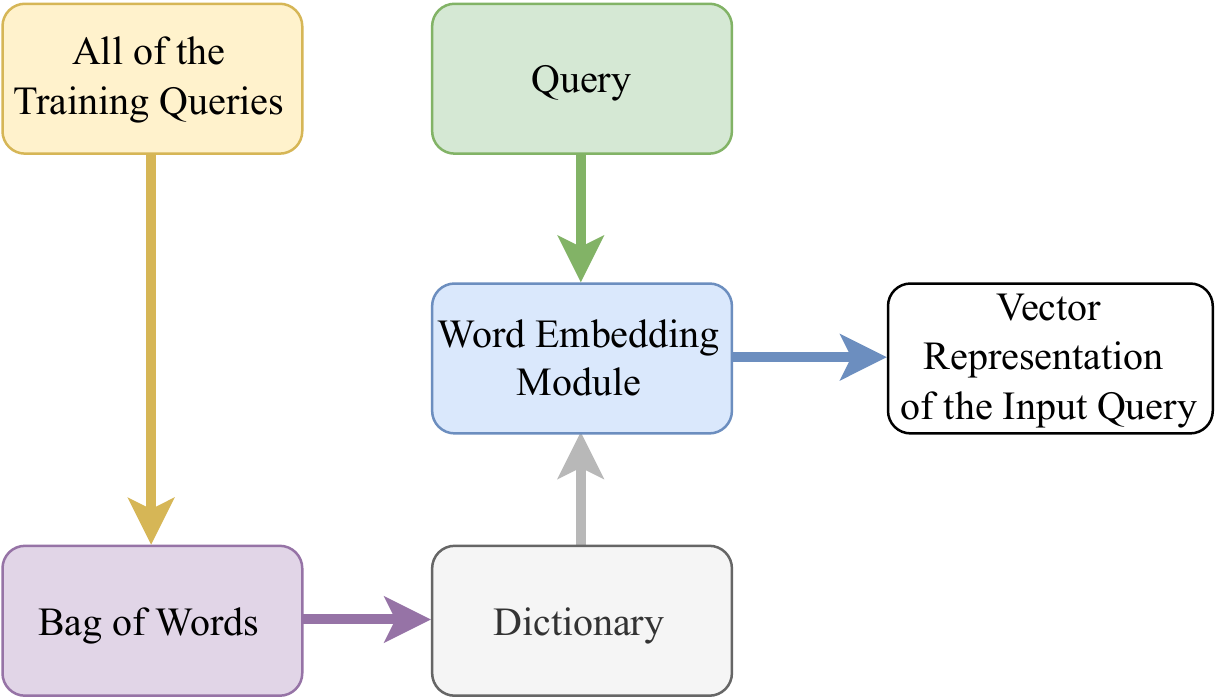}
\end{center}
\vspace{-0.40cm}
   \caption{This figure conceptually depicts our video summary controller. In the video summary controller, we take all the queries from our training set to form a bag of words and create a dictionary. Then, we base on this dictionary to embed an input query.}
\vspace{-0.30cm}
\label{fig:figure6}
\end{figure}

% \vspace{+0.10cm}
\noindent\textbf{3.2 Crowd-sourced Annotation}

In this subsection, we analyze the frame-based relevance score annotations obtained through the above procedure. Also, we explain how we merge these relevance score annotations for each video into one set of ground truth labels.

\noindent\textbf{Label distributions of relevance scores.}
The distribution of relevance score annotations is ``Very Good'': 18.65\%, ``Good'': 55.33\%, ``Not good'': 13.03\% and ``Bad'': 12.99\%.
% Training set - 0:2806, 1:3035, 2:12348, 3:4497
% Validation set - 0:1333, 1:927, 2:3738, 3:1564
% Testing set - 0:772, 1:968, 2:4834, 3:988
% Sum:        - , 4911, 4930, 20920, 7049
% Total (Train+Val+Test): 37810

% \textcolor{red}{Relevance annotation consistency: Given the inherent subjectivity of the task, we want to know whether annotators agree with each other about the query relevance of frames. To do this, we follow previous work \cite{gygli2013interestingness,isola2011understanding,wang2016event} and compute the Spearman rank correlation (rho) between the relevance scores of different subjects, splitting five annotations of each video into two groups of two and three raters each. We take all split combination to find mean (rho) for a video. Our dataset has an average correlation of rho = 0.73 over all videos, where 1 is a perfect correlation while 0 would indicate no consistency in the scores. On the related task of event-specific image importance, using five annotators, consistency is only rho = 0.4 \cite{wang2016event}. Thus, we can be confident that our relevance labels are of high quality. \textcolor{blue}{maybe ignore this part.}}

\noindent\textbf{Ground truth.}
As mentioned in the \textit{``Video Summarization Dataset Comparison''} subsection, human-based evaluation is problematic and time-consuming \cite{gygli2014creating}. So, in this work, based on \cite{vasudevan2017query}, the following approach are used for evaluation. For evaluation of the testing videos, one way is to ask human experts to watch the full video, instead of just video summaries, and access the relevance of every single part of the video. Then, their responses are considered as gold standard annotations \cite{gygli2014creating,khosla2013large,potapov2014category}. The advantage of this approach is that once the annotations are obtained, experiments can be carried out indefinitely. This is desirable, especially for a computer vision system involving multiple iterations and testing. Note that, in the proposed dataset, we create a single ground truth relevance score label for each query-video pair by merging the corresponding relevance score annotations from AMT workers. Then, we base on the majority vote rule, \cite{antol2015vqa,huang2019novel}, to evaluate the model performance for a relevance score prediction, i.e., a predicted relevance score is correct if the majority of human annotators provided that exact score. Note that we map annotations, ``Very Good'' to 3, ``Good'' to 2, ``Not Good'' to 1, and ``Bad'' to 0.
Note that, referring to \ref{table:table103}, the relevance score in this work and the importance score from TVSum \cite{song2015tvsum} are different. The relevance score in this work is to capture the relation between a given text-based query and a video frame. The important score is to capture the importance between a video frame and a final video summary of the video \cite{song2015tvsum}.

\section{Method}

\noindent\textbf{Overview.}
In this section, we start to describe the proposed query-controllable video summarization method. The proposed method is composed of a video summary controller, video summary generator, and video summary output module. The summary controller takes a text-based query as input and outputs the vector representation of the query. The summary generator takes the embedded query and a video as inputs and outputs the frame-based relevance score prediction. Finally, the video summary output module will use the score prediction to generate a video summary. Figure \ref{fig:figure100} explains the above procedure.

% \vspace{+0.37cm}
\noindent\textbf{4.1 Video Summary Controller}

Text-based queries are meant to represent the expected video summary content while subtly alludes its semantic relationship. Therefore, we use the following way to encode input queries and add their contribution to our proposed method. In our paper, we exploit the vector representation of the text-based input query to control the generated video summary. The main idea of our video summary controller is to generate a vector representation of an input query, based on a dictionary. In the beginning, we form a dictionary based on a bag of words which are collected from all the unique words of the training queries. Then, we encode an input query by exploiting the dictionary. After the encoding, we have a vector representation of the input query to represent the expected video summary content. To make the procedure clearer, we make a flowchart to explain the above, referring to Figure \ref{fig:figure6}.

\begin{figure*}
  \includegraphics[width=\textwidth]{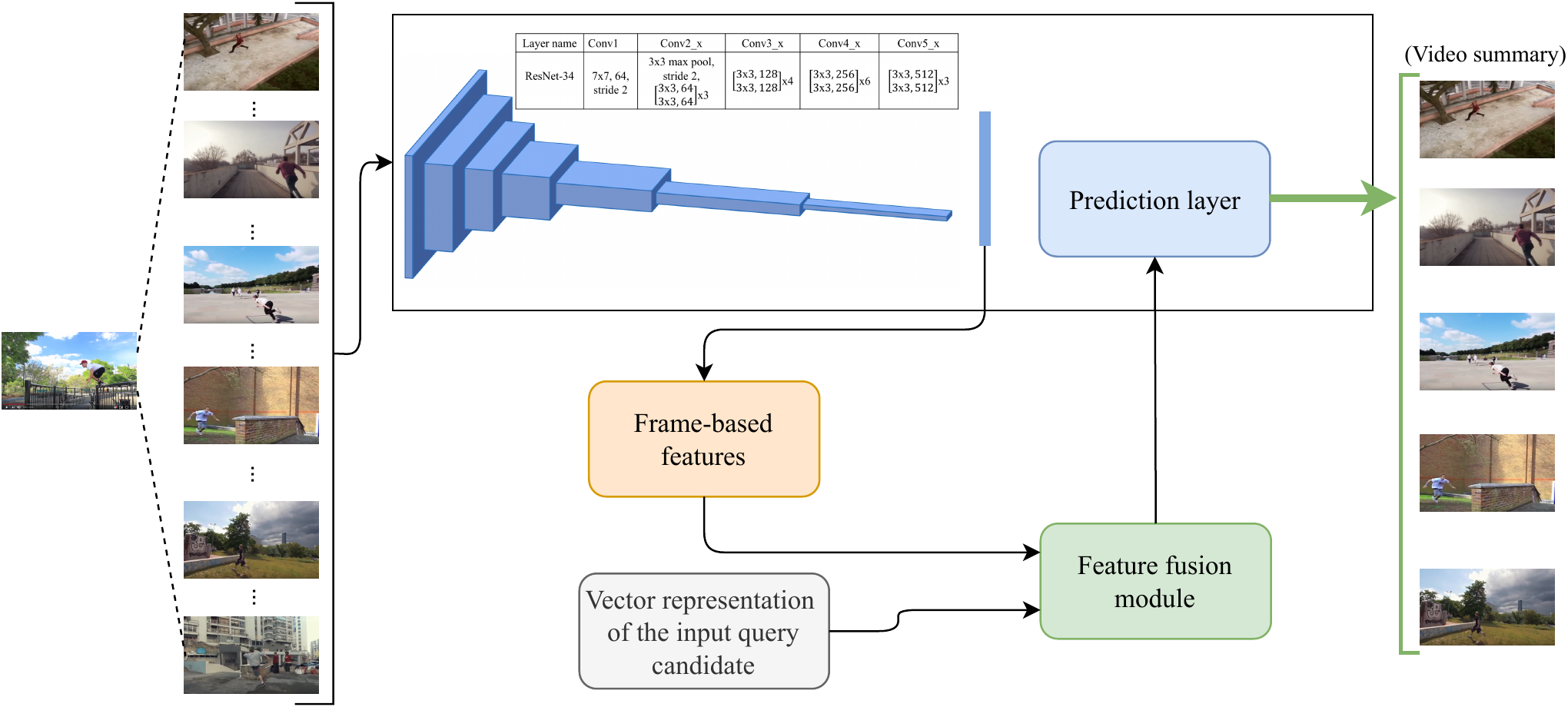}
  \vspace{-0.8cm}
  \caption{
    This figure conceptually depicts our video summary generator. In the generator, we take an input video and sample it at 1 fps. Then, we input the sampled frames to a CNN-based structure for extracting frame-based features. The feature will be fused with a text-based input query feature. Finally, the fused feature will be pass to a prediction layer to generate the frame-based relevance scores. Finally, we base on the predicted relevance scores to output a query-dependent video summary.}
%   \Description{Enjoying the baseball game from the third-base seats. Ichiro Suzuki preparing to bat.}
  \label{fig:figure1}
  \vspace{-0.3cm}
\end{figure*}

\noindent\textbf{4.2 Video Summary Generator}

The main idea of the video summary generator is to take a vector representation of an input text-based query and a video to generate a frame-based relevance score vector. The summary generator is composed of a convolutional neural network (CNN) structure and a multi-modality features fusion module. Note that the CNN structure will be trained on our training set. Before an input video goes to the CNN structure, it is sampled at 1 fps. Then, in our case, we use ResNet-34, \cite{he2016deep}, to extract the 199 frame-based features for each input video. Note that the feature used is the visual layer one layer below the classification layer. After the features are extracted, we exploit a feature fusion module to fuse the frames-based features and the input text-based query feature. The fused feature vector will be sent to a fully connected layer for the frame-based relevance score prediction. The feature fusion module will be depicted in the following subsection. Please refer to Figure \ref{fig:figure1} for the flowchart of the above procedure. Note that we take ``Cross-Entropy Loss'' as our loss function, referring to Equation \ref{eq:loss}, and Adam \cite{kingma2014adam} as our optimizer. For the optimizer parameters, coefficients used for computing moving averages of gradient and its square are $\beta_{1}=0.9$ and $\beta_{1}=0.999$, respectively. The term added to the denominator to improve numerical stability is $\epsilon=1e-8$, and the learning rate $\alpha=1e-4$.

\begin{equation}
\vspace{-0.2cm}
    \label{eq:loss}
    Loss(x, class) = -x[class]+ln(\sum_{j}exp(x[j])),
\end{equation}

where ``$class$'' denotes the ground truth class, and ``$x$'' indicates the prediction.

% There are some existing DNN visual explanation methods, such as \cite{zhou2015cnnlocalization,selvaraju2017grad}. The authors of \cite{zhou2015cnnlocalization} have proposed a technique, called Class Activation Mapping (CAM), for CNN. It makes classification-trained CNN learn how to perform the task of object localization, without using a bounding box. Furthermore, they exploit class activation maps to visualize the predicted class scores on a given image, highlighting the discriminative object parts which are detected by the CNN. To improve the conventional retinal diseases treatment procedure, we incorporate the DNN visual explanation module in our proposed method. Also, we exploit this module to help us analyze the effectiveness of the method, referring to EFFECTIVENESS EVALUATION OF THE PROPOSED METHOD section for more details.

\noindent\textbf{Multi-modality Features Fusion Module.}
One of the technical problems of our proposed method is fusing the query and frame-based features with minimum loss of information. It has been shown that, in similar multi-modal contexts, the performance of models can decrease if models are poorly designed; how to solve this issue, in general, remains an open question \cite{ben2017mutan,fukui2016multimodal}. In this work, we exploit 3 difference commonly used approaches, summation, concatenation, and element-wise multiplication, to fuse query and frame-based features.

% \vspace{+0.10cm}
\noindent\textbf{4.3 Video Summary Output Module}

After we get the frame-based relevance score prediction vector from the video summary generator, we pass this vector to our video summary output module. The main idea of this module is to output a video summary based on the relevance score prediction vector. In our case, we map labels, “Very Good” to 3, “Good” to 2, “Not Good” to 1, and“Bad” to 0. If a predicted relevance score greater than or equal to 2, then we consider the corresponding frame is relevant. If a predicted relevance score is less than 2, then we consider the corresponding frame is irrelevant. Finally, we collect $k$ relevant frames in time order as our video summary. Note that $k$ is a user-defined parameter for the length of a video summary.

\section{Experiments and Analysis}
In this section, we will evaluate our proposed end-to-end method for the query-controllable video summarization task based on the setup of the proposed dataset. We will also analyze the effectiveness of the query and methods of multi-modal features fusion.

\noindent\textbf{5.1 Dataset preparation}

To validate our proposed query-controllable video summarization method, we base on the following dataset setup to conduct our experiments. We separate the whole dataset into 60\%/20\%/20\%, i.e., 114/38/38, for training/validation/testing, respectively. One video has one corresponding query. The maximum number of words of the query is 8. Regarding the frame size input of CNN is 128 by 128 with 3 channels, i.e., red, green, and blue. Note that we normalize each image channel by $mean = (0.4280, 0.4106, 0.3589)$ and $std = (0.2737, 0.2631, 0.2601)$. The maximum number of frames of video is 199. Similar to the video preprocessing method in \cite{sigurdsson2017asynchronous}, we make all the videos have the same number of frames, i.e., 199. We show the original number of frames of each video in Figure \ref{fig:figure2}.

% \vspace{+0.10cm}
\noindent\textbf{5.2 Effectiveness Analysis of Query}

\begin{figure}[t!]
\begin{center}
\includegraphics[width=1.0\linewidth]{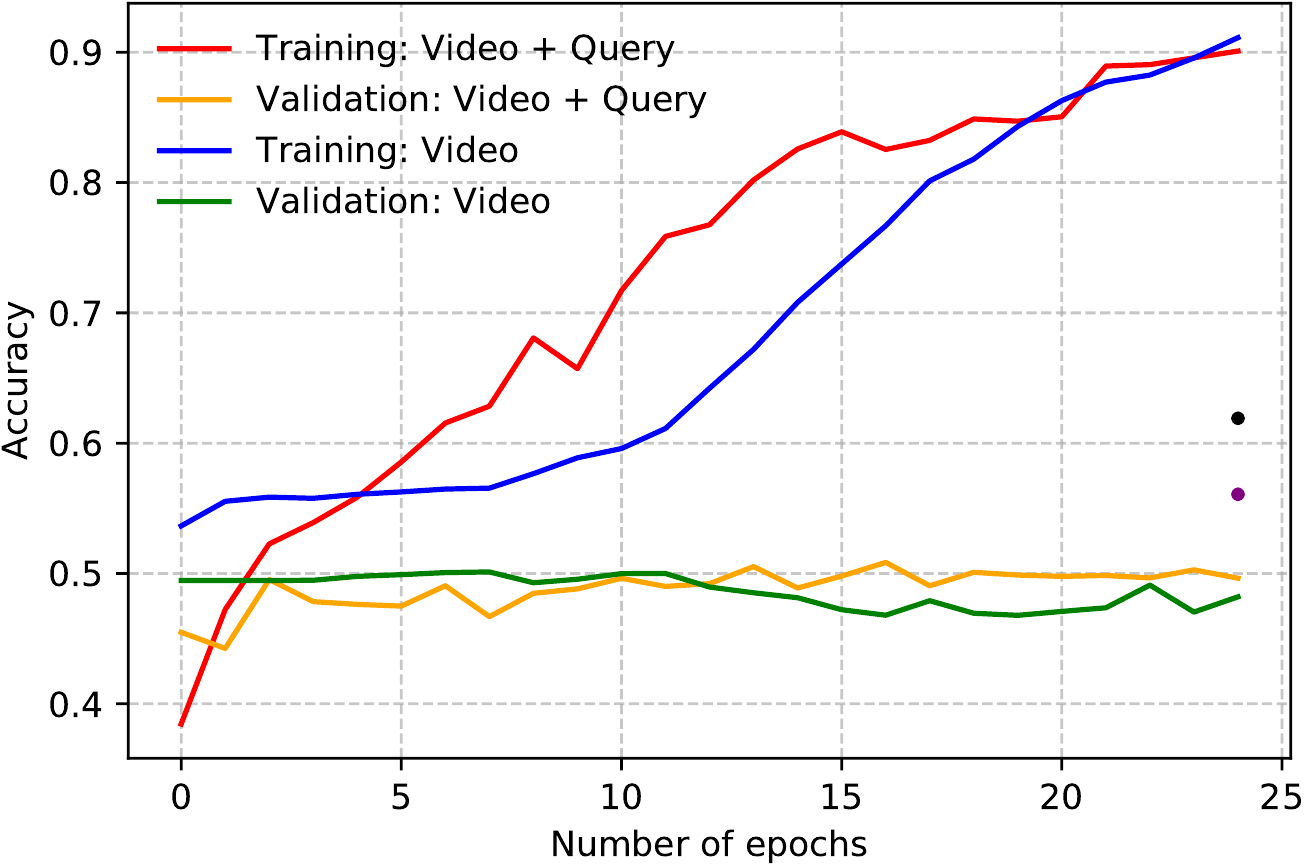}
\end{center}
\vspace{-0.40cm}
   \caption{The figure shows the model performance in two cases, ``video only'' and ``video+query''. In this figure, the x-axis denotes the number of epochs. The y-axis denotes the model accuracy. We test models after the 25 epochs of training. Note that the ``purple point'' denotes the ``video only'' testing accuracy, 0.5608, and the ``black point'' indicates the ``video+query'' testing accuracy, 0.6191.}
\vspace{-0.30cm}
\label{fig:figure5}
\end{figure}

\begin{figure}[h!]
\begin{center}
\includegraphics[width=1.0\linewidth]{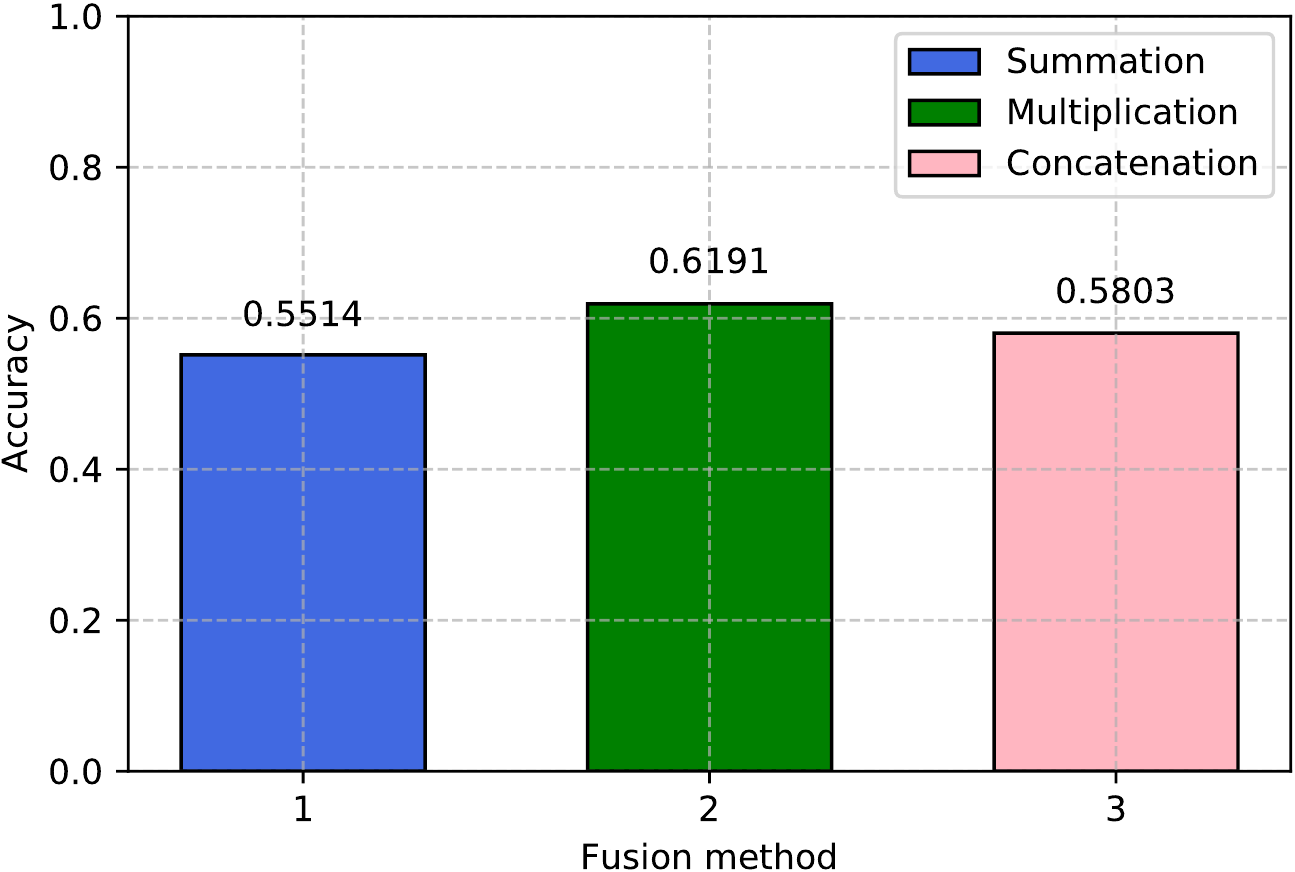}
\end{center}
\vspace{-0.40cm}
   \caption{The figure shows the model performance under three different cases, i.e., multi-modal features fusion by ``summation'', ``concatenation'', and the ``element-wise multiplication''. The model with element-wise multiplication fusion method has the best performance. In this figure, the x-axis denotes the case index. The y-axis denotes the model accuracy. Note that each model is well trained with a different number of epochs.}
\vspace{-0.30cm}
\label{fig:figure41}
\end{figure}

In this experiment, we want to know whether the text-based query will help generate a better video summary or not. So, we conduct the experiment based on two types of models, query-driven, and non-query-driven. According to Figure \ref{fig:figure5}, we discover that the query-driven model, with the testing accuracy 0.6191, is better than the non-query-driven one, with the testing accuracy 0.5608. Also, based on the validation accuracy versus the number of epochs, we can see that the textual query is capable of guiding the query-driven model to perform better the non-query-driven one. However, when we compare the worst model, fusing features by summation, in Figure \ref{fig:figure41} to the non-query-driven model, we discover that the non-query-driven model performs better. This motivates us to conduct the other experiment about the comparison of multi-modal feature fusion methods, referring to the next subsection.

% \begin{figure}[h!]
% \begin{center}
% \includegraphics[width=1.0\linewidth]{query_test.pdf}
% \end{center}
% % \vspace{-0.30cm}
%   \caption{The figure shows the model performance in two cases, ``video only'' and ``video+query''. In this figure, the x-axis denotes the case index. The y-axis denotes the model accuracy. Note that each model is well trained with a different number of epochs.}
% % \vspace{-0.30cm}
% \label{fig:figure5}
% \end{figure}

% \vspace{+0.10cm}
\noindent\textbf{5.3 Effectiveness Analysis of Different Fusion Methods}

In general, multi-modal  features fusion is still an open question \cite{ben2017mutan,fukui2016multimodal}. So, we base on the commonly used methods, summation, concatenation, and the element-wise multiplication, to conduct our experiment. According to Figure \ref{fig:figure41}, we can see that the model with element-wise multiplication fusion method has the best performance. If we compare the performance of the other 2 feature fusion methods to the non-query-driven model, we discover that the model with the concatenation fusion method is still better than the non-query-driven model. However, the model with the summation fusion method is worse than the non-query-driven model.

\noindent\textbf{Interaction between query and video.} Based on the result of Figure \ref{fig:figure5} and Figure \ref{fig:figure41}, it implies that using a proper multi-modal  feature fusion method is important. The reason is that the fused feature embeds the implicit interaction between video, i.e., frames, and query. If we use an improper fusion method, such as summation, the query will confuse the network in some sense. This situation also happens in \cite{antol2015vqa}.

% \vspace{+0.10cm}
\noindent\textbf{5.4 Qualitative Results and Analysis}

\begin{figure*}[!tbp]
  \begin{subfigure}[b]{0.9\textwidth}
    \includegraphics[width=0.98\linewidth]{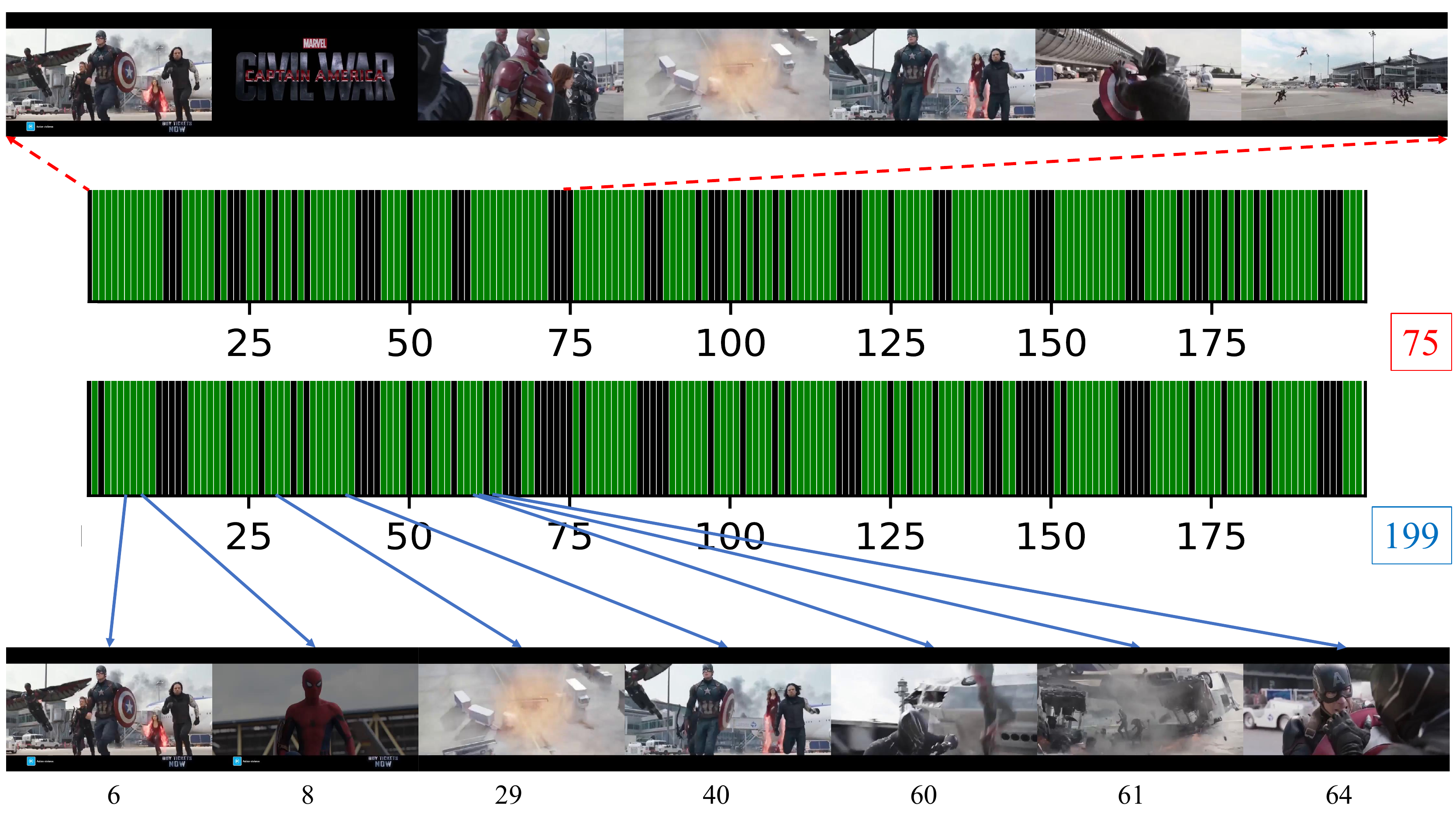}
    \caption{Query: ``civil war spiderman''. Correct number of relevance score prediction $/$ Total number of frames = $94/199$.}
    %  \vspace{+10pt}
  \end{subfigure}
  \hfill
  \begin{subfigure}[b]{0.9\textwidth}
    \includegraphics[width=0.98\linewidth]{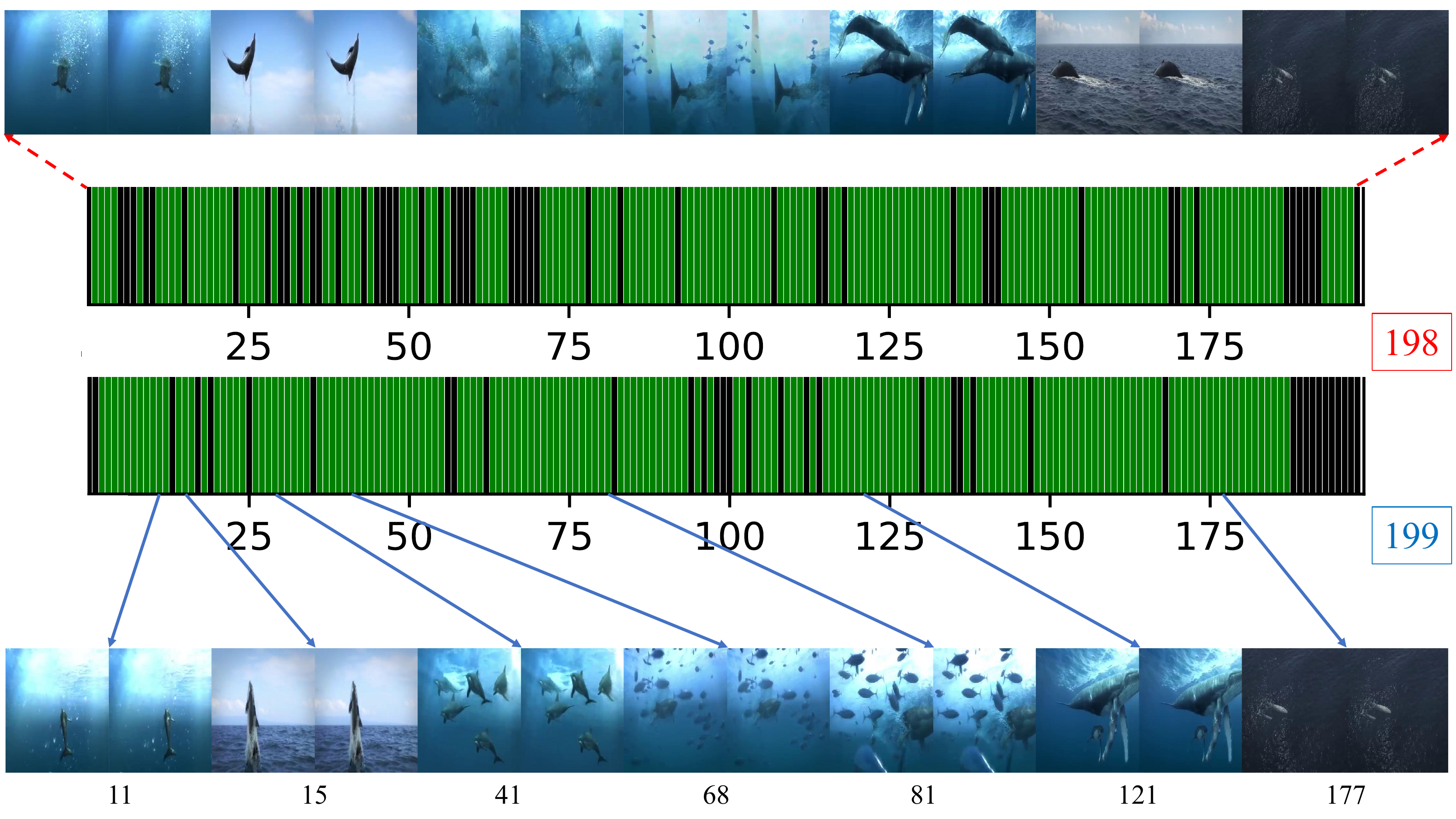}
    \caption{Query: ``3d movies''. Correct number of relevance score prediction $/$ Total number of frames = $120/199$.}
  \end{subfigure}
  \caption{In this figure, based on a similar qualitative result visualization method from \cite{rochan2018video}, we show two generated video summaries with the corresponding queries. In (a), ``75'' color-coded by red indicates the original total number of frames of the input video. In the first row, we show some frames from the original video, with ``75'' frames, to represent the input video. The second row represents the original input video. The third row represents the prediction. In the fourth row, we show $k$, e.g., $k=7$, frames to represent our generated video summary. The numbers at the bottom indicate the frame index in the original video. Note that the video frame index is starting from 0. In (b), we show the second generated video summary result and the corresponding notations are similar to (a).}
\label{fig:figure4}
% \vspace{-3pt}
\end{figure*}

In this subsection, we show some qualitative results, illustrated in Figure \ref{fig:figure4}. Note that because of the limited space, we are only able to show some frames to represent the original video and the corresponding generated video summary in time order. In Figure \ref{fig:figure4}-(a), the input is a video with the query ``civil war spiderman''. Based on our \textit{``Video Summary Output Module''} subsection, we use ``green'' to indicate the relevant frame and ``black'' to indicate irrelevant frames. The second row in (a) represents the video frames with ground truth labels. The third row in (a) represents the video frames with predicted labels. The correct number of relevance score prediction is 94 out of 199. Note that the number of frames of the original input video is 75, so we only show the video summary frames selected from 0 to 74 in order. In Figure \ref{fig:figure4}-(b), the input is another video with the query ``3d movies''. We also use the same color to indicate the relevant and irrelevant frames. The second row of (b) indicates the video frames with ground truth labels.  The third row in (b) represents the video frames with predicted labels. The correct number of relevance score prediction is 120 out of 199. Similar to (a), since the number of frames of the original input video is 198, so we only show the video summary frames selected from 0 to 197 in order. Based on Figure \ref{fig:figure4}, it shows that our proposed method is capable of generating video summaries with content relevant to the input query.

% \textcolor{red}{According to our experimental results in Table \ref{table:table1}, we find that the established DNNs-based models outperform the U-Net-SVM method. Although the results in \cite{yang2018novel} shows that U-Net-SVM model outperforms the DNNs-based models in the small dataset, DNNs-based models have better performance in the practical scenario for clinical applications with the large-scale dataset, referring to Table \ref{table:table1}. Note that ResNet50 model with the pretrained weights of ImageNet has the best performance in our case. Also, based on Table \ref{table:table1}, it is worth noticing that ResNet50, VGG16, VGG19 outperform deeper models like DenseNet121, DenseNet161, InceptionV3, ResNet101, and ResNet152. We conjecture that the models, more complex than ResNet50, tend to overfit our proposed DEN dataset while ResNet50 can learn more robust features. Since the DNN-based module achieves the human-level performance, apparently it makes the proposed method effective and also successfully improves the traditional retinal diseases treatment procedure.} \textcolor{blue}{All models are trained on the training set for $200$ epochs.}

\section{Conclusion and Future Work}

To sum up, we treat a query-controllable video summarization task as a supervised learning problem in this work. To tackle this problem, we propose an end-to-end deep learning based approach to generate a query-dependent video summary. The proposed method contains a video summary controller, video summary generator, and video summary output module. To foster the query-controllable video summarization research and conduct our experiments, we propose a new dataset. Each video in the proposed dataset is annotated by frame-based relevance score labels. Our experimental results show that the text-based query not only helps control video summary, but also improves the model performance with +$5.83$\% in the sense of accuracy. Based on our experiment, we know that the multi-modal feature fusion method is crucial, so developing a new fusion approach will be interesting future work.

\begin{acks}
This project has received funding from the European Union’s Horizon 2020 research and innovation programme under the Marie Skłodowska-Curie grant agreement No 765140.
\end{acks}

%
% The next two lines define the bibliography style to be used, and the bibliography file.
\bibliographystyle{ACM-Reference-Format}
\bibliography{sample-base}

% 
% If your work has an appendix, this is the place to put it.
\appendix



% Lorem ipsum dolor sit amet, consectetur adipiscing elit. Morbi malesuada, quam in pulvinar varius, metus nunc fermentum urna, id sollicitudin purus odio sit amet enim. Aliquam ullamcorper eu ipsum vel mollis. Curabitur quis dictum nisl. Phasellus vel semper risus, et lacinia dolor. Integer ultricies commodo sem nec semper. 


% Etiam commodo feugiat nisl pulvinar pellentesque. Etiam auctor sodales ligula, non varius nibh pulvinar semper. Suspendisse nec lectus non ipsum convallis congue hendrerit vitae sapien. Donec at laoreet eros. Vivamus non purus placerat, scelerisque diam eu, cursus ante. Etiam aliquam tortor auctor efficitur mattis. 


% Nam id fermentum dui. Suspendisse sagittis tortor a nulla mollis, in pulvinar ex pretium. Sed interdum orci quis metus euismod, et sagittis enim maximus. Vestibulum gravida massa ut felis suscipit congue. Quisque mattis elit a risus ultrices commodo venenatis eget dui. Etiam sagittis eleifend elementum. 

% Nam interdum magna at lectus dignissim, ac dignissim lorem rhoncus. Maecenas eu arcu ac neque placerat aliquam. Nunc pulvinar massa et mattis lacinia.

\end{document}